\documentclass[%
aps
 amsmath,amssymb,
preprint,%
]{revtex4-1}
\usepackage{accents}
\usepackage{subfigure}
\usepackage{graphicx}% Include figure files
\usepackage{dcolumn}% Align table columns on decimal point
\usepackage{bm}% bold math
\usepackage{appendix}
\usepackage[dvipsnames]{xcolor}
\begin{document}

%\preprint{APS/123-QED}

\title{Anomalous edge plasma transport, neutrals, and divertor plasma detachment}% Force line breaks with \\

\author{Yanzeng Zhang}
\author{Sergei I. Krasheninnikov}
\author{Rebecca Masline}
\author{Roman D. Smirnov}
\affiliation{%
 $^1$Mechanical and Aerospace Engineering Department, University of California San Diego, La Jolla, CA 92093, USA
}

\begin{abstract}

An impact of neutrals on anomalous edge plasma transport and zonal flow (ZF) is considered. As an example, it is assumed that edge plasma turbulence is driven by the resistive drift wave (RDW) instability. It is found that the actual effect of neutrals is not related to a suppression of the instability \textit{per se}, but due to an impact on the ZF. Particularly, it is shown that, whereas the neutrals make very little impact on the linear growth rate of the RDW instability, they can largely reduce the zonal flow generation in the nonlinear stage, which results in an enhancement of the overall anomalous plasma transport. Even though only RDW instability is considered, it seems that such an impact of neutrals on anomalous edge plasma transport has a very generic feature. It is conceivable that such neutral induced enhancement of anomalous plasma transport is observed experimentally in a detached divertor regime, which is accompanied by a strong increase of neutral density.

\end{abstract}

%\keyworZD{Suggested keyworZD}%Use showkeys class option if keyword
\maketitle

\section{Introduction}
Due to plasma neutralization on plasma facing components and volumetric recombination processes, edge plasma in magnetic fusion devices contains significant amount of neutrals. These neutrals play crucial role in both plasma fueling (resulting in so-called plasma ``recycling'') and in establishing the regime of divertor plasma detachment \cite{krasheninnikov2017physics}. In addition, there is a significant amount of papers, both experimental \cite{tsuchiya1996effect,pedrosa1995influence,zweben2014effect,horton2000h,doyle2007plasma} and theoretical \cite{connor2000review,daughton1998interchange,odblom1999neutrals,monier1997effects,fulop2001effect,d2002effect}, discussing the role of neutrals in plasma anomalous cross-field transport and in the transition from L-mode to H-mode confinement regime. 

Whereas experimental data demonstrate some controversial/indecisive conclusions with regard to neutral effects in edge plasma turbulence (e.g. see \cite{horton2000h,doyle2007plasma} and the references therein), early, most analytical and theoretical estimates were showing that the role of neutrals could be important. Recently, the results of numerical simulations of edge plasma turbulence have demonstrated that the incorporation of neutral effects is causing significant and, in some cases, very strong impact \cite{wersal2015first,stotler2017neutral,thrysoe2018plasma,bisai2018influence}. Although, to deduce physics behind this impact additional post processing of 3D plasma turbulence simulation results should be performed, which in most cases is not available.

Potentially strong impact of neutrals on the processes governing plasma turbulence could be seen from simplified fluid neutral model, which is based on a strong coupling of plasma ions and neutral atoms caused by ion-neutral collisions (e.g. see \cite{helander1994fluid} and the references therein). Indeed, in this case the contribution of neutrals, which are not magnetized, to combined plasma/neutral cross-field viscosity coefficient, $\eta_N$, is given by the following expression
\begin{equation}
    \eta_N\approx \frac{N}{n}\frac{T_i/M}{\nu_{Ni}},\label{eq_eta_n}
\end{equation}
where $N$ and $n$ are the neutral and plasma density, $T_i$ is the ion temperature, $M$ is the ion mass, $\nu_{Ni}=K_{Ni}\cdot n$ is the neutral-ion collision frequency, and $K_{Ni}$ is the ion-neutral collision rate constant. For $T_i\approx 10eV$, $K_{Ni}\approx 3\cdot 10^{-8}cm^3/s$, and $n\approx 3\cdot 10^{13}cm^{-3}$, from Eq. (1) we find $\eta_N\approx 10^7 (N/n)[cm^2/s]$. As a result, for rather typical edge plasma ratio $N/n\tilde{>} 10^{-3}$, $\eta_N$ exceeds characteristic anomalous edge plasma diffusivities $\sim 10^4[cm^2/s]$. 

Such simplified fluid description of plasma-neutral coupling allowing to study the neutral impact on plasma dynamics (e.g., plasma instability) only holds for the case where: i) elastic neutral-ion collisions dominate over the rate of electron impact ionization of neutrals, and ii) the characteristic spatial scale (e.g., wavelength $\lambda$) and the frequency, $\omega$, of the problem under consideration are, respectively, larger and smaller than the mean-free path of neutrals with respect to neutral-ion collisions, $\lambda_{Ni}=\sqrt{T_i/M}/\nu_{Ni}$, and $\nu_{Ni}$. 

However, in practice, the situation is more complex. First of all, edge plasma parameters vary strongly and whereas in divertor region of current tokamaks plasma temperature could be below $10 eV$ and plasma density could exceed $10^{14} cm^{-3}$, in the vicinity of the separatrix at the midplane of the scrape-off layer (SOL) temperature could reach $100 eV$ and density could be below $10^{13} cm^{-3}$ and falls even more at the main chamber wall. This shows that the fluid approximation for atomic hydrogen transport is not valid in the entire edge plasma volume. Next, in addition to neutral-ion elastic collisions, neutrals undergo the electron impact ionization process accompanied by a significant amount of the radiation loss. For electron temperature above $ \sim  10 eV$, the neutral ionization rate constant becomes comparable with the rate constant of elastic (charge-exchange) collisions of neutral atoms with plasma protons/deuterons (e.g. see \cite{krasheninnikov2017physics} and the references therein). Finally, apart from hydrogen atoms, a large fraction ($ \sim  50\%$ ) of hydrogen comes from the walls as molecules, which have a much lower rate of elastic collisions with plasma ions (protons/deuterons). Therefore, a fluid approximation for hydrogen molecule transport is even more problematic than that for hydrogen atoms. For the case where either of inequalities, $\lambda_{Ni}<\lambda$, $\omega<\nu_{Ni}$, needed to ensure the application of the fluid neutral model is violated, taking into account that elastic neutral-ion collisions in a ballpark dominate, one could consider neutrals as stationary ``scatters''  of ions, causing ion momentum loss. 

As a result, rigorous consideration of the impact of neutrals on the linear stage of edge plasma instabilities becomes a rather complex problem. On the other hand, the most important practical issue is the impact of neutrals on anomalous edge plasma transport, associated with the nonlinear stage of instabilities. However, the nonlinear stage is strongly impacted by the generation of plasma zonal flows (ZF) (e.g., see \cite{diamond2005zonal,fujisawa2008review} and the references therein). Therefore, it is plausible that the actual effect of neutrals on edge plasma anomalous transport is not related to the suppression of instability \textit{per se}, but due to an impact on ZF, which has a strong impact on both nonlinear stage of instability and anomalous transport. In this case, the most crucial part is an appropriate description of an impact of neutrals on ZF.

In this paper, we examine the impact of neutrals on resistive drift-wave (RDW) turbulence and plasma transport described with modified \cite{numata2007bifurcation} Hasegawa-Wakatani \cite{hasegawa1983plasma} (MHW) equations. We find that indeed, whereas neutrals make very little impact on the growth rate of the RDW instability, they have a pronounced effect on ZF and, therefore, on overall anomalous plasma transport.

We notice that the transition to the regime of divertor plasma detachment results in a strong increase of neutral density in the divertor volume \cite{krasheninnikov2017physics}. Therefore, it is conceivable the neutral density at the core-edge interface is also increasing during the transition into detachment. It could reduce the amplitude of ZF and provoke an enhancement of anomalous cross-field plasma transport. Interestingly, the correlation between the divertor plasma detachment and the increase of cross-field plasma transport, causing the broadening of the width of the scrape-off layer, was observed in experiments \cite{sun2015study}. To see how strongly neutral density at the core-edge interface increases in the course of the transition into divertor plasma detachment, we performed corresponding simulations of the edge plasma transport with the code UEDGE \cite{rognlien1992fully}. We have found that indeed neutral density just inside the separatrix is strongly increased after the transition into the detached regime and, therefore, could significantly alter the ZF, and thus the anomalous plasma transport.

The rest of the paper is organized as follows. In Section \ref{sec_analysis} we present the results of our analytic consideration of an impact of neutrals on both RDW instability and ZF generation. In section \ref{sec_numerical} we discuss the results of our numerical simulation of an impact of neutrals on the RDW turbulence (including ZFs) performed within the framework of modified Hasegawa-Wakatani (MHW) model. In section \ref{sec_detachment} we present the results of our simulations of an impact of the transition into detached divertor regime on neutral density at the core-edge interface. The main results are summarized in section \ref{sec_conclusion}.

\section{Impact of neutrals on RDW instability and ZF generation}
\label{sec_analysis}
We consider the RDW instability and ZF generation assuming cold ions and constant electron temperature. To account for an impact of neutrals on ion dynamics we add the term $\mathbf{S}_{neut}=M\hat{s}_{neut}\mathbf{V}_i$ into ion equation of motion:
\begin{equation}
    M\frac{d\mathbf{V}_i}{dt}=-e\nabla \phi +\frac{e}{c}(\mathbf{V}_i\times \mathbf{B})-\mathbf{S}_{neut},\label{eq_ion_moment}
\end{equation}
where the operator $\hat{s}_{neut}$ is defined as
\begin{equation}
\hat{s}_{neut}=\left\{
             \begin{array}{lr}
             -\eta_N\nabla^2, \textup{for fluid model}  \\
             (N/n)\nu_{Ni}, \textup{for ``scattering'' neutrals}&  
             \end{array}
\right.,\label{eq_definition_hat_s_neut}
\end{equation}
$\mathbf{B}=\mathbf{e}_zB_0$ is the magnetic field, $\phi$ is the fluctuating electrostatic potential, and $\mathbf{V}_i$ is the ion velocity. Then by employing the standard drift-orders from Eq.~(\ref{eq_ion_moment}), we obtain a modification of the MHW model to account for the neutral impact on the RDW and ZF
\begin{eqnarray}
&&\rho_s^2\left[\frac{\partial}{\partial t} +\mathbf{V}\cdot \nabla+\hat{s}_{neut} \right]\nabla^2\frac{e\phi}{T_e}=\nu_\parallel \left(\frac{e\tilde{\phi}}{T_e}-\frac{\tilde{\mathcal{N}}}{n_0}\right)+\nu\rho_s^2\nabla^2 \nabla^2\frac{e\phi}{T_e},\label{eq_vorticity}\\
&&\frac{\partial}{\partial t} \frac{\mathcal{N}}{n_0}+\mathbf{V}\cdot \nabla \frac{\mathcal{N}}{n_0}=\nu_\parallel \left(\frac{e\tilde{\phi}}{T_e}-\frac{\tilde{\mathcal{N}}}{n_0}\right)-V_*\frac{\partial }{\partial y}\frac{e\phi}{T_e}+D\nabla^2 \frac{\mathcal{N}}{n_0},\label{eq_density}
\end{eqnarray}
where $\rho_s^2=C_s^2/\Omega_{Bi}^2$, $C_s^2=T_e/M$,  $  \Omega _{Bi}=eB_0/cM$, $\mathbf{V}=c\mathbf{e}_z\times\nabla\phi/B_0$, $V_*=cT_e/eB_0L_n$, $L_{n}^{-1}=-d\ell n({{n}_{0}})/dx$, $T_e$ is electron temperature, $\mathcal{N}$ is the plasma density fluctuations, $\nu_\parallel=k_z^2T_e/(m\nu_{ei})$, $m$ is electron mass, and  $\nu _{ei}$  is electron collision frequency.
Here the resistive coupling term between the electrostatic potential and plasma density fluctuations is determined only by the non-zonal components $\tilde{f}=f-\left<f\right>_y$, where $\left<f\right>_y\equiv \int_0^{L_y} f dy/L_y$ denotes the integration along the poloidal line at a given radial location. The last terms on the right hand side of Eqs.~(\ref{eq_vorticity}, \ref{eq_density}) with constant coefficients $D$ and $\nu$ are dissipation terms for the purpose of numerical stability and will be ignored in the following analysis. 

We first consider the neutral impact on the RDW by deriving its dispersion equation. For this purpose, we ignore the ZF and adopt the eikonal approximation (and thus $\nabla^2$ in fluid $\hat{s}_{neut}$ is replaced by $-k_\perp^2$). Moreover, we assume $\nu_\parallel$ to be larger than both the RDW frequency $\omega$ and $\omega_*\equiv k_yV_*$. As a result, we arrive to the following dispersion equation
\begin{equation}
    \frac{\omega_*}{\omega}=1+\rho_s^2k_\perp^2+i\left(\rho_s^2k_\perp^2\frac{\hat{s}_{neut}}{\omega}+\frac{\omega-\omega_*}{\nu_\parallel}\right).\label{eq_dispersion}
\end{equation}
From Eq.~(\ref{eq_dispersion}) we find that an impact of neutrals stabilizes the RDW instability for 
\begin{equation}
    \hat{s}_{neut}>\hat{s}_{th}^{dw}\equiv\omega_*^2/\nu_\parallel(1+\rho_s^2k_\perp^2)^2,\label{eq_thresho_suppress_instab}
\end{equation}
where $\omega\approx \omega_*/(1+\rho_s^2k_\perp^2)$. Assuming that $\rho_s^2k_\perp^2\sim 1$, for $\omega_*/\nu_\parallel\sim 0.4$ and $\omega_*\sim 3\times 10^5s^{-1}$, this inequality shows that even for ``scattering''  neutral model the RDW is stabilized for the neutral density $N\tilde{>}10^{12}cm^{-3}$, which is too high for the SOL plasma in most of tokamaks and, therefore, the stabilization of the RDW instability is only possible in divertor volume where neutral density is significantly higher.

Next, we examine an impact of neutrals on the generation of ZF by averaging Eq.~(\ref{eq_vorticity}) over the poloidal direction. As a result, we obtain the evolution equation for the mean flow
\begin{equation}
    \frac{\partial}{\partial t}\left<\phi\right>_y=\frac{c}{B_0}\left<\frac{\partial \tilde{\phi}}{\partial x}\frac{\partial \tilde{\phi}}{\partial y}\right>_y-\hat{s}_{neut}^{zf}\left<\phi\right>_y.\label{eq_evolution_zf}
\end{equation}
Note that the large-scale mean flow varies on a longer time scale compared to the small-scale RDW fluctuations and we assume that the mean flow is one dimensional so that $\nabla_y\to 0$, while the small-scale fluctuations are two dimensional $\tilde{\phi}=\tilde{\phi}(x,y)$. As a result, for the fluid model, $\hat{s}_{neut}$ defined in Eq.~(\ref{eq_definition_hat_s_neut}) should be averaged along the poloidal direction and we have $\hat{s}_{neut}^{zf}=-\eta_Nd^2/dx^2$, whereas, for the scattering neutrals, $\hat{s}_{neut}^{zf}=\hat{s}_{neut}=(N/n)\nu_{Ni}$.

To quantify $\left<\partial_x \tilde{\phi}\partial_y\tilde{\phi}\right>_y$, considering that Eqs.~(\ref{eq_vorticity}, \ref{eq_density}) reduce to the modified Hasegawa-Mima \cite{hasegawa1978pseudo} equations in the adiabatic limit, $\nu_\parallel> \omega,~\omega_*$, if we ignore the contribution of neutrals to the RDW when $\hat{s}_{neut}< \omega_*^2/\nu_\parallel$ as shown in Eq.~(\ref{eq_thresho_suppress_instab}), it can be conveniently computed from the kinetic equation for the drift wave action \cite{smolyakov2000coherent}. As a result, if we assume that the ZF fluctuations are described by the frequency $\Omega< \omega$ and radial wave number $q=-id/dx$ (and thus $d/dx^2$ in the fluid description of $\hat{s}_{neut}^{zf}$ is replaced by $-q^2$), which is assumed to be smaller than the width of the drift wave spectrum $N_0^k=(1+\rho_s^2k_\perp^2)|e\tilde{\phi}_k/T_e|^2$  in $k_x$, we find 
\begin{equation}
    -i\Omega=-\left(\frac{c}{B_0}\right)^2q^2\int\frac{R(\Omega,q,\Delta \omega_k)k_y^2k_x}{(1+\rho_s^2k_\perp^2)^2}\frac{\partial N_0^k}{\partial x}d\mathbf{k}_\perp-\hat{s}_{neut}^{zf},\label{eq_dispersion_ZF}
\end{equation}
where $R(\Omega,q,\Delta \omega_k)=i/(\Omega-qV_g+i\Delta\omega_k)$ is the response function with $\Delta\omega_k$ being the nonlinear broadening increment and $V_g=\partial \omega/\partial k_x$. Eq.~(\ref{eq_dispersion_ZF}) shows that the neutrals cause suppression of the ZF generation. 

Here we consider a generation of the ZF resulting from a monochromatic spectrum $N_0^k=N_0\delta(\mathbf{k}_\perp-\mathbf{k}_0)$, where the growth rate of ZF generation is larger compared to the resonant instability corresponding to a broad spectrum of drift waves \cite{smolyakov2000zonal}. This assumption is consistent with the simulation observation in the next section, where $\rho_sk_{0y}\approx 1$ and $\rho_sk_{0x}< 1$. As a result, from Eq.~(\ref{eq_dispersion_ZF}) we obtain 
\begin{equation}
    1+\frac{\sigma^2}{(\Omega-qV_g)^2}=-i\frac{\hat{s}_{neut}^{zf}}{\Omega},\label{eq_supression_ZF}
\end{equation}
where $\sigma=qC_s\rho_s|k_{0y}|N_0^{1/2}(1+\rho_s^2k_0^2-4\rho_s^2k_{0x}^2)^{1/2}(1+\rho_s^2k_0^2)^{-3/2}$ characterizes the growth rate of ZF in the absence of neutrals. Given that $V_g\propto \rho_sk_{0x}< 1$, we can ignore $qV_g$ in Eq.~(\ref{eq_supression_ZF}). As a result, we obtain the growth rate of ZF:
\begin{equation}
    \gamma_{zf}=\sqrt{\sigma^2+(\hat{s}_{neut}^{zf})^2/4}-\hat{s}_{neut}^{zf}/2.\label{eq_growth_zf}
\end{equation}
It follows that the neutrals can largely reduce the ZF growth rate when
\begin{equation}
\hat{s}_{neut}^{zf}>\hat{s}_{th}^{zf}\equiv\sigma\sim \kappa\rho_s q \omega_*,  \label{eq_suppss_ZF_final}
\end{equation}
where we take $N_0^{1/2}\sim|e\tilde{\phi}/T_e|\approx\kappa/k_yL_n$, and $\kappa< 1$ is a factor for weak fluctuations before nonlinear saturation. From Eqs.~(\ref{eq_thresho_suppress_instab}, \ref{eq_suppss_ZF_final}) we see that the impact of neutrals on the ZF generation is stronger than on the RDW instability when $\kappa \rho_sq< \omega_*/\nu_\parallel$ ($\hat{s}_{th}^{zf}< \hat{s}_{th}^{dw}$) and thus it is plausible that the neutrals will enhance the RDW turbulence by reducing the ZF rather than affect the RDW instability. From the numerical simulations in the next section we will see that the neutrals with small density ($N< 10^{11}cm^{-3}$) can indeed reduce the ZF intensity resulting in an enhancement of the RDW turbulence transport, but this occurs mainly in the nonlinear regime while in the linear state the impact of neutrals is negligible as $\hat{s}_{neut}<\hat{s}_{th}^{dw}\sim \hat{s}_{th}^{zf}$.

So far we consider an impact of neutrals on both development of the RDW turbulence and ZF generation and neglect an impact of ions. Such approximation is justified for the case where $\hat{s}_{neut}>\hat{s}_{ion}$, here $\hat{s}_{ion}$ is the analog of $\hat{s}_{neut}$ caused by ions. Assuming that both $\rho_ik_\perp$ and $q\rho_i$ are below unity, we conclude that $\hat{s}_{ion}$ is due to cross-field ion viscosity, whereas $\hat{s}_{neut}$ could be determined by neutrals in a “scattering” regime. In the latter case, using the expression for cross-field ion viscosity~\cite{braginskii1965transport} and the ion-neutral collision frequency from \cite{helander1994fluid}, we find that for $\rho_i^2k_\perp^2\sim q^2\rho_i^2\sim 1$ and ion temperature $\sim 100 eV$, $\hat{s}_{neut}>\hat{s}_{ion}$ for $N/n>10^{-4}$. In our further considerations we assume that the latter inequality holds.

\section{Numerical simulation of an impact of neutrals on RDW turbulence and plasma transport}
\label{sec_numerical}

In this section, Eqs.~(\ref{eq_vorticity}, \ref{eq_density}) will be numerically solved to examine the impact of neutrals on the RDW turbulence and plasma transport. The numerical scheme used is a pseudo-spectral Fourier code by employing Dedalus\cite{2019arXiv190510388B}, where the computation domain is a square box with size $L_x = L_y = 20\pi\rho_s$ so that the lowest wavenumber is $\rho_s\Delta k = 0.2$. We employ the doubly periodic boundary conditions and the number of the modes are chosen as $256\times 256$. The fourth-order Runge-Kutta method is chosen as the time integration algorithm with a time step $\Delta t=10^{-3}\rho_s/V_*$, where $\rho_s/V_*\sim \omega_*^{-1}$ is the characteristic temporal scale of RDW turbulence. In fact, we can rescale $e\phi/T_e$ and $\mathcal{N}/n_0$ by $V_*/C_s=\rho_s/L_n$ and, as a result, the system is controlled by two free parameters $\alpha=\nu_\parallel \rho_s/V_*$ and $\hat{s}=\hat{s}_{neut}\rho_s/V_*$ (here we consider the ``scattering'' neutrals only and thus $\hat{s}_{neut}=\hat{s}_{neut}^{zf}$ is a constant coefficient), where the dissipation coefficients $\nu= D=10^{-4}V_*\rho_s$ are fixed. In the simulations, we also fix $\alpha=2$ but vary $\hat{s}=(0, 1,~10)\times 10^{-3}$ corresponding to $N\sim (0,~1,~10)\times 10^{10} cm^{-3}$ for $V_*/\rho_s=3\times 10^5s^{-1}$. We are particularly interested in the impact of the neutrals on the normalized particle flux in the radial direction
\begin{equation}
    \Gamma_n\equiv -\int \left(\frac{L_n}{\rho_s}\frac{\tilde{\mathcal{N}}}{n_0}\right)\times\rho_s\frac{\partial}{\partial y}\left(\frac{L_n}{\rho_s}\frac{e\tilde{\phi}}{T_e}\right)d\mathbf{x},\label{eq_particle_flux}
\end{equation}
where $\int fd\mathbf{x}\equiv\int_0^{L_x}\int_0^{L_y}fdxdy/L_xL_y$.

\begin{figure}[bht]
\centering
\subfigure{
\begin{minipage}{0.45\textwidth}
\includegraphics[width=1\textwidth]{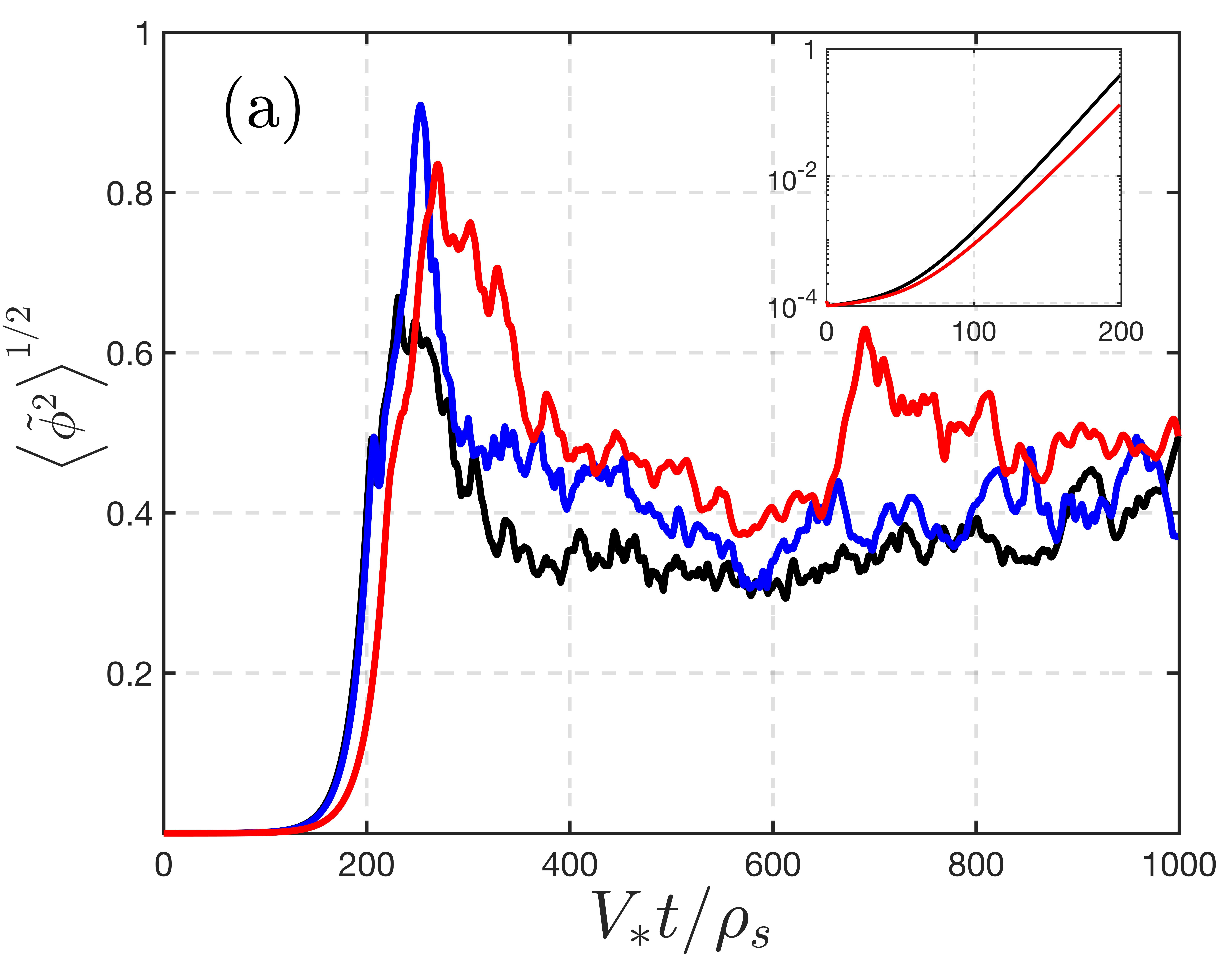}
\label{fig-fluctuation-average-phi}
\end{minipage}}
\subfigure{
\begin{minipage}{0.45\textwidth}
\includegraphics[width=1\textwidth]{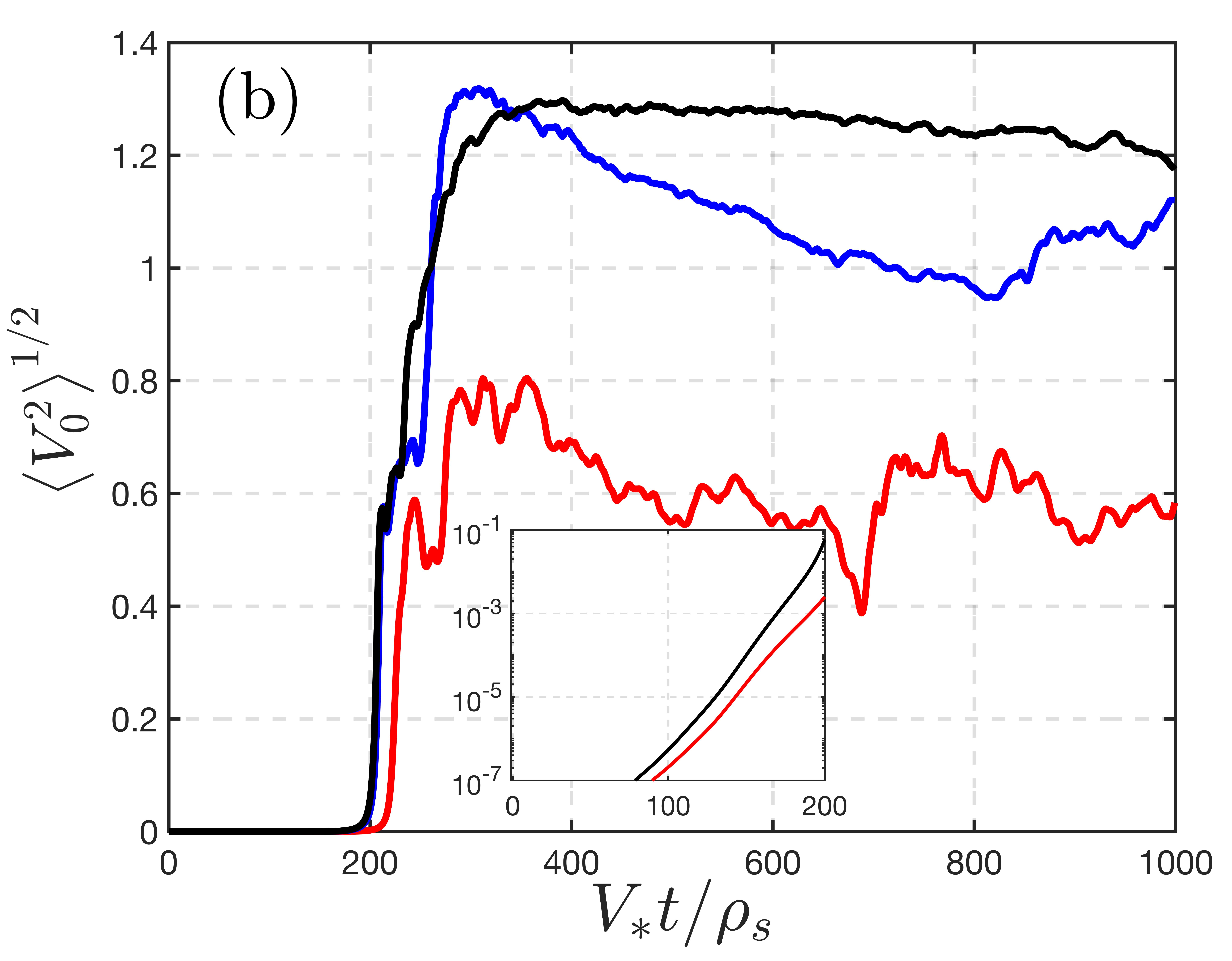}
\label{fig-fluctuation-average-ZF}
\end{minipage}}
\caption{Time evolution plots of root mean square of the normalized (a) electrostatic potential fluctuations and (b) zonal flow for $\alpha=2$, where the insets plot results in the linear stage in the logarithmic scale. The black, blue and red curves are for $\hat{s}=0,~10^{-3}$ and $10^{-2}$.}
\label{fig-fluctuation-average}
\end{figure}

In all the cases, we start the simulations from the same small amplitude perturbation with $\mathcal{N}/n_0,e\phi/T_e\propto exp(-\mathbf{x}^2/\Delta^2)$, where $\Delta= 4\rho_s$. We observe that, the perturbations will first grow due to the RDW instability, e.g., see Fig.~\ref{fig-fluctuation-average-phi} for the root mean square (RMS) of the normalized electrostatic potential fluctuations $\left<\tilde{\phi}^2\right>^{1/2}\equiv[\int (L_ne\tilde{\phi}/\rho_sT_e)^2d\mathbf{x}]^{1/2}$. In the late stage of the linear regime, the dominant mode of $\rho_s k_y\approx 1$ and $\rho_sk_x\approx 0.2$ dominates and thus the fluctuations grow linearly as shown in the inset plot, where the growth rate from the simulations matches the analyses. Particularly, without neutrals, the growth rate calculated from both the numerical simulations and analyses are $\gamma_0\approx k_y^2 V_*^2/\nu_\parallel(1+\rho_s^2k_\perp^2)^3\approx 0.059 V_*/\rho_s$. Whereas, taking into account $\hat{s}_{th}^{dw}\rho_s/V_*=0.12$ from Eq.~(\ref{eq_thresho_suppress_instab}), the neutrals even with $\hat{s}=10^{-2}$ can only reduce the growth rate $\gamma_0$ by approximately $10\%$, which agrees with the numerical simulations. Meanwhile, these neutrals reduce the linear growth rate of the ZF by $\sim 20\%$ as shown in the inset plot of Fig.~\ref{fig-fluctuation-average-ZF}, indicating $\hat{s}_{neut}/\sigma\sim 0.4$ from Eq.~(\ref{eq_growth_zf}). This is consistent with the estimate in Eq.~(\ref{eq_suppss_ZF_final}) for $\kappa=L_ne\tilde{\phi}/T_e\rho_s \sim 0.1$ as shown in Fig.~\ref{fig-fluctuation-average-phi}, where $q\rho_s\sim 0.4$ in all the cases.

\begin{figure}[bht]
\centering
\begin{minipage}{0.6\textwidth}
\includegraphics[width=1\textwidth]{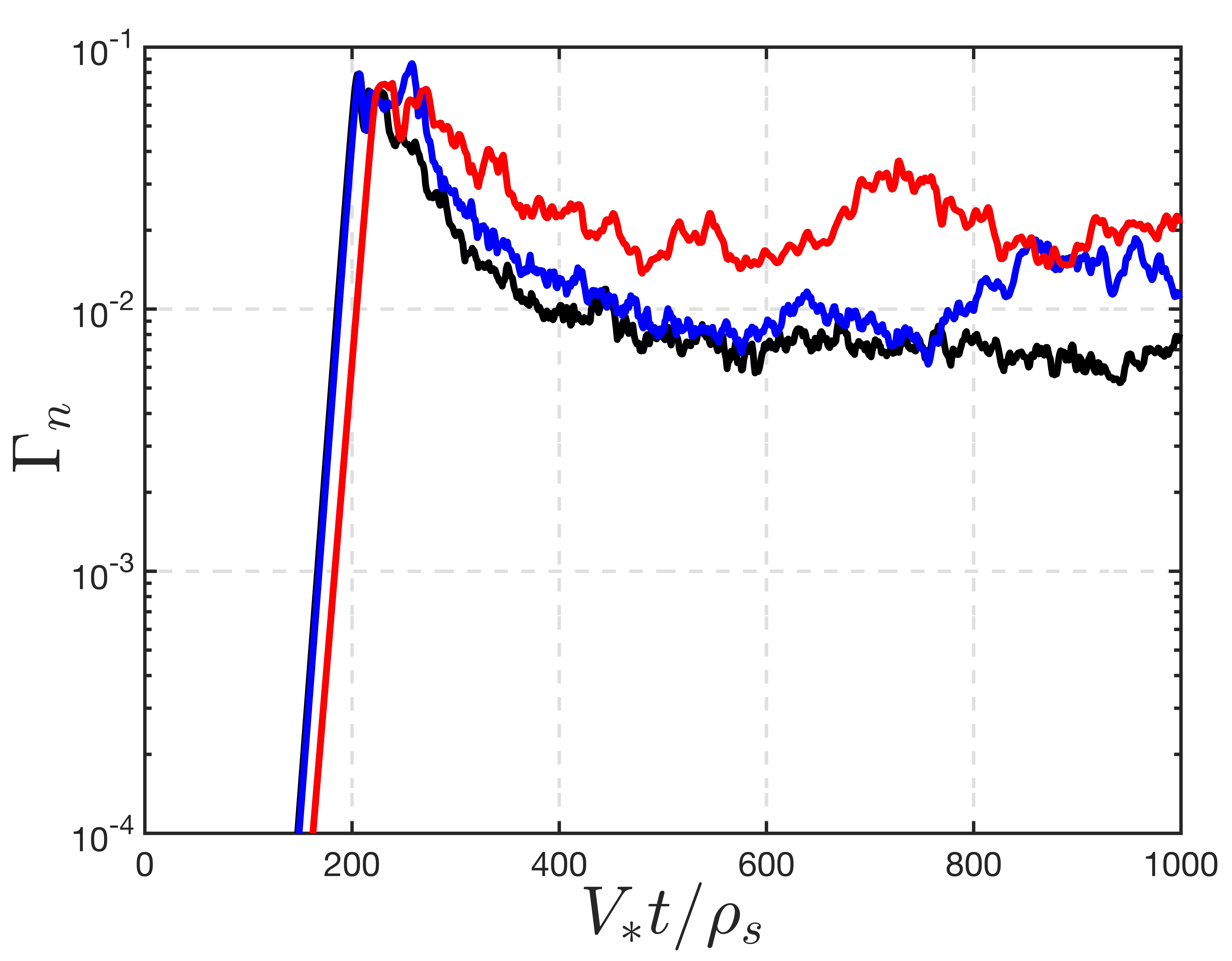}
\end{minipage}
\caption{Time evolution plots of the normalized particle flux for $\alpha=2$, where the black, blue and red curves are for $\hat{s}=0,~10^{-3}$ and $10^{-2}$.}
\label{fig-particle_flux}
\end{figure}

However, the neutrals play an important role in the nonlinear stage when the system is saturated (it's worthy noting that, in the presence of large neutral concentrations, ZF-RDW system undergoes large predator–prey oscillations\cite{diamond1994self}, but they are on a larger time scale than the RDW). As we can see from Fig.~\ref{fig-fluctuation-average}, the ZFs are largely reduced by the neutrals while the fluctuation level is increased (the physics underlying the reduction of ZF in the nonlinear regime will be considered somewhere else). As a result, the transport of particle flux is enhanced as shown in Fig.~\ref{fig-particle_flux}, where the time-averaged $\Gamma_n$ at the saturated state from $V_*t/\rho_s=400$ to $V_*\rho_s=1000$ are $\Gamma_n=7.4\times10^{-3},~ 1.1\times10^{-2}$ and $2.0\times10^{-2}$ for $\hat{s}=0,~10^{-3}$ and $10^{-2}$, respectively. 

\section{Neutral density at the core-edge interface during the transition into detached divertor plasma}
\label{sec_detachment}

To address the issue of neutral density variation at the core-edge interface during the transition into detached divertor regime we perform a series of the UEDGE simulations of edge plasma for DIII-D-like geometry and magnetic configuration (see~\cite{masline2019influence} for details). Self-consistent modeling of plasma detachment in a high confinement mode (H-mode) of the operation of a tokamak and an impact of neutrals on anomalous cross-field plasma transport goes beyond the scope of this paper. Therefore, here we just illustrate an impact of the transition into detached divertor regime on neutral density at the core-edge interface assuming that cross-field plasma transport coefficients are fixed and equal to $1 m^2/s$. It is known that the transition to divertor plasma detachment could be reached by the increase of impurity fraction or plasma density, or both \cite{krasheninnikov2017physics}. In our simulations we take fixed plasma density at the core-edge interface equal to $2\times 10^{13}cm^{-3}$, assume constant power flux of $4 MW$ into edge plasma from the core (split evenly between electron and ion components), and gradually increase nitrogen impurity fraction in the simulation domain (for simplicity we used so-called ``fixed impurity fraction model'').  We monitor detachment process by the magnitude of plasma flux onto divertor targets, $\Gamma_{\textup{div}}$, since one of the signature of divertor plasma detachment is the reduction of $\Gamma_{\textup{div}}$~\cite{krasheninnikov2017physics}.

\begin{figure}[bht]
\centering
\begin{minipage}{0.65\textwidth}
\includegraphics[width=1\textwidth]{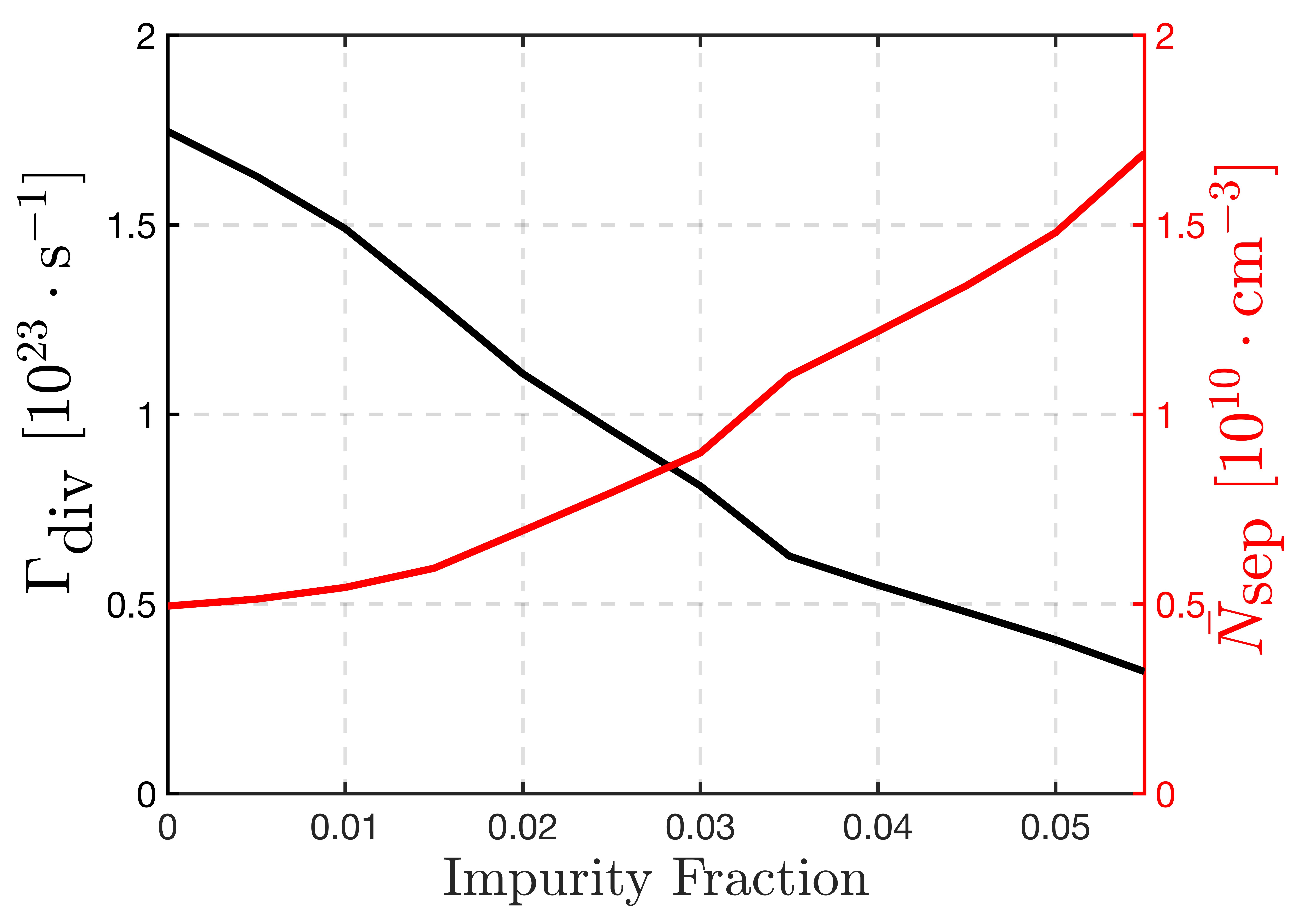}
\label{fig-detached-flux}
\end{minipage}
\caption{Plasma flux to the divertor targets (black) and neutral density just inside the separatrix (red) as the functions of impurity fraction.}
\label{fig-detached-plasma}
\end{figure}

In Fig.~\ref{fig-detached-plasma} we present both $\Gamma_{\textup{div}}$ and the neutral density averaged over the flux tube just inside the separatrix, $\bar{N}_{\textup{sep}}$, as the functions of impurity fraction. One can clearly see that the transition into divertor detachment, manifested as the reduction of $\Gamma_{\textup{div}}$, is accompanied by a strong increase of $\bar{N}_{\textup{sep}}$. There are two reasons accounting for this increase of neutral density: i) an overall increase of neutral density in a low temperature detached divertor plasma and ii) the expansion of this region (detachment front) toward the X-point\cite{krasheninnikov2017physics}, which reduces an effective opacity of the scrape-off layer plasma for neutral penetration through the separatrix.

We notice that the neutral density exceeding  $10^{10}cm^{-3}$ found in the most detached case in our series of runs, corresponds to the parameter $\hat{s}>10^{-3}$, and, according to out simulations of the RDW turbulence (recall Fig.~\ref{fig-particle_flux}), can make a profound impact on ZF and anomalous plasma transport. We also notice that, in our UEDGE simulations, we consider the simplest model of edge plasma transport. More complete/complicated models show that the transition to detachment can have the bifurcation-like character \cite{krasheninnikov2017stability,jaervinen2018b}, which could be accompanied by corresponding bifurcation of neutral density inside the separatrix and anomalous transport.

\section{Conclusions}
\label{sec_conclusion}

In this paper, the impact of neutrals on the RDW and turbulence transport is examined for the tokamak edge plasmas by employing the MHW model, which characterizes the basic physics of the RDW-ZF system. The neutrals effect on the ion dynamics are modeled both in the fluid regime, when the wavelength and frequency of the RDW are, respectively, larger and smaller than the mean-free path of neutrals with respect to neutral-ion collisions and the collision frequency, and the ``scattering'' regime for the opposite conditions. From the analysis, we find that the neutrals make very little impact on the RDW instability in the linear regime, which agrees well with the numerical simulations. This is also true for the linear ZF, where the fraction of the reduced growth rate due to the neutrals is comparable to that of the RDW. However, the neutrals can largely reduce the ZF in the nonlinear stage as shown in Fig.~\ref{fig-fluctuation-average-ZF}, which, in turn, leads to an enhancement of the radial particle flux in Fig.~\ref{fig-particle_flux}. Even though only RDW instability is considered, it seems that such an impact of neutrals on anomalous edge plasma transport has very generic feature (e.g., the neutrals impact on the ITG turbulence was simulated in \cite{stotler2017neutral}). 

Our numerical simulations of the transition of edge plasma into detached divertor regime with the UEDGE transport code show that such transition is accompanied by a strong increase of neutral density at the core-edge interface, which, according to our turbulence simulations, results in significant reduction of the amplitude of ZF and an increase of anomalous cross-field plasma transport. This could explain experimental data from \cite{sun2015study}, showing the widening of the scrape-off layer width after transition into detached divertor regime. 

However, we notice that in our model the neutrals are simplified as a uniform background affecting only the ion momentum balance. Therefore, an ultimate conclusion of the role of neutrals in the anomalous cross-field edge plasma transport and the transition to plasma detachment requires a more comprehensive self-consistent evolution of neutrals, which is beyond the scope of this paper.

\section*{ACKNOWLEDGMENTS}

This work was supported by the U.S. Department of Energy, Office of Science, Office of Fusion Energy Sciences under Award No. DE-FG02-04ER54739 at UCSD.

\bibliography{main}% Produces the bibliography via BibTeX.

\begin{thebibliography}{32}%
\makeatletter
\providecommand \@ifxundefined [1]{%
 \@ifx{#1\undefined}
}%
\providecommand \@ifnum [1]{%
 \ifnum #1\expandafter \@firstoftwo
 \else \expandafter \@secondoftwo
 \fi
}%
\providecommand \@ifx [1]{%
 \ifx #1\expandafter \@firstoftwo
 \else \expandafter \@secondoftwo
 \fi
}%
\providecommand \natexlab [1]{#1}%
\providecommand \enquote  [1]{``#1''}%
\providecommand \bibnamefont  [1]{#1}%
\providecommand \bibfnamefont [1]{#1}%
\providecommand \citenamefont [1]{#1}%
\providecommand \href@noop [0]{\@secondoftwo}%
\providecommand \href [0]{\begingroup \@sanitize@url \@href}%
\providecommand \@href[1]{\@@startlink{#1}\@@href}%
\providecommand \@@href[1]{\endgroup#1\@@endlink}%
\providecommand \@sanitize@url [0]{\catcode `\\12\catcode `\$12\catcode
  `\&12\catcode `\#12\catcode `\^12\catcode `\_12\catcode `\%12\relax}%
\providecommand \@@startlink[1]{}%
\providecommand \@@endlink[0]{}%
\providecommand \url  [0]{\begingroup\@sanitize@url \@url }%
\providecommand \@url [1]{\endgroup\@href {#1}{\urlprefix }}%
\providecommand \urlprefix  [0]{URL }%
\providecommand \Eprint [0]{\href }%
\providecommand \doibase [0]{http://dx.doi.org/}%
\providecommand \selectlanguage [0]{\@gobble}%
\providecommand \bibinfo  [0]{\@secondoftwo}%
\providecommand \bibfield  [0]{\@secondoftwo}%
\providecommand \translation [1]{[#1]}%
\providecommand \BibitemOpen [0]{}%
\providecommand \bibitemStop [0]{}%
\providecommand \bibitemNoStop [0]{.\EOS\space}%
\providecommand \EOS [0]{\spacefactor3000\relax}%
\providecommand \BibitemShut  [1]{\csname bibitem#1\endcsname}%
\let\auto@bib@innerbib\@empty
%</preamble>
\bibitem [{\citenamefont {Krasheninnikov}\ and\ \citenamefont
  {Kukushkin}(2017)}]{krasheninnikov2017physics}%
  \BibitemOpen
  \bibfield  {author} {\bibinfo {author} {\bibfnamefont {S.}~\bibnamefont
  {Krasheninnikov}}\ and\ \bibinfo {author} {\bibfnamefont {A.}~\bibnamefont
  {Kukushkin}},\ }\href@noop {} {\bibfield  {journal} {\bibinfo  {journal}
  {Journal of Plasma Physics}\ }\textbf {\bibinfo {volume} {83}} (\bibinfo
  {year} {2017})}\BibitemShut {NoStop}%
\bibitem [{\citenamefont {Tsuchiya}\ \emph {et~al.}(1996)\citenamefont
  {Tsuchiya}, \citenamefont {Takenaga}, \citenamefont {Fukuda}, \citenamefont
  {Kamada}, \citenamefont {Ishida}, \citenamefont {Sato}, \citenamefont
  {Takizuka} \emph {et~al.}}]{tsuchiya1996effect}%
  \BibitemOpen
  \bibfield  {author} {\bibinfo {author} {\bibfnamefont {K.}~\bibnamefont
  {Tsuchiya}}, \bibinfo {author} {\bibfnamefont {H.}~\bibnamefont {Takenaga}},
  \bibinfo {author} {\bibfnamefont {T.}~\bibnamefont {Fukuda}}, \bibinfo
  {author} {\bibfnamefont {Y.}~\bibnamefont {Kamada}}, \bibinfo {author}
  {\bibfnamefont {S.}~\bibnamefont {Ishida}}, \bibinfo {author} {\bibfnamefont
  {M.}~\bibnamefont {Sato}}, \bibinfo {author} {\bibfnamefont {T.}~\bibnamefont
  {Takizuka}},  \emph {et~al.},\ }\href@noop {} {\bibfield  {journal} {\bibinfo
   {journal} {Plasma Physics and Controlled Fusion}\ }\textbf {\bibinfo
  {volume} {38}},\ \bibinfo {pages} {1295} (\bibinfo {year}
  {1996})}\BibitemShut {NoStop}%
\bibitem [{\citenamefont {Pedrosa}\ \emph {et~al.}(1995)\citenamefont
  {Pedrosa}, \citenamefont {Garc{\'\i}a-Cort{\'e}s}, \citenamefont {Branas},
  \citenamefont {Balbin}, \citenamefont {Hidalgo}, \citenamefont {Schmitz},
  \citenamefont {Tynan},\ and\ \citenamefont
  {Post-Zwicker}}]{pedrosa1995influence}%
  \BibitemOpen
  \bibfield  {author} {\bibinfo {author} {\bibfnamefont {M.}~\bibnamefont
  {Pedrosa}}, \bibinfo {author} {\bibfnamefont {I.}~\bibnamefont
  {Garc{\'\i}a-Cort{\'e}s}}, \bibinfo {author} {\bibfnamefont {B.}~\bibnamefont
  {Branas}}, \bibinfo {author} {\bibfnamefont {R.}~\bibnamefont {Balbin}},
  \bibinfo {author} {\bibfnamefont {C.}~\bibnamefont {Hidalgo}}, \bibinfo
  {author} {\bibfnamefont {L.}~\bibnamefont {Schmitz}}, \bibinfo {author}
  {\bibfnamefont {G.}~\bibnamefont {Tynan}}, \ and\ \bibinfo {author}
  {\bibfnamefont {A.}~\bibnamefont {Post-Zwicker}},\ }\href@noop {} {\bibfield
  {journal} {\bibinfo  {journal} {Physics of Plasmas}\ }\textbf {\bibinfo
  {volume} {2}},\ \bibinfo {pages} {2618} (\bibinfo {year} {1995})}\BibitemShut
  {NoStop}%
\bibitem [{\citenamefont {Zweben}\ \emph {et~al.}(2014)\citenamefont {Zweben},
  \citenamefont {Stotler}, \citenamefont {Bell}, \citenamefont {Davis},
  \citenamefont {Kaye}, \citenamefont {LeBlanc}, \citenamefont {Maqueda},
  \citenamefont {Meier}, \citenamefont {Munsat}, \citenamefont {Ren} \emph
  {et~al.}}]{zweben2014effect}%
  \BibitemOpen
  \bibfield  {author} {\bibinfo {author} {\bibfnamefont {S.}~\bibnamefont
  {Zweben}}, \bibinfo {author} {\bibfnamefont {D.}~\bibnamefont {Stotler}},
  \bibinfo {author} {\bibfnamefont {R.}~\bibnamefont {Bell}}, \bibinfo {author}
  {\bibfnamefont {W.}~\bibnamefont {Davis}}, \bibinfo {author} {\bibfnamefont
  {S.}~\bibnamefont {Kaye}}, \bibinfo {author} {\bibfnamefont {B.}~\bibnamefont
  {LeBlanc}}, \bibinfo {author} {\bibfnamefont {R.}~\bibnamefont {Maqueda}},
  \bibinfo {author} {\bibfnamefont {E.}~\bibnamefont {Meier}}, \bibinfo
  {author} {\bibfnamefont {T.}~\bibnamefont {Munsat}}, \bibinfo {author}
  {\bibfnamefont {Y.}~\bibnamefont {Ren}},  \emph {et~al.},\ }\href@noop {}
  {\bibfield  {journal} {\bibinfo  {journal} {Plasma Physics and Controlled
  Fusion}\ }\textbf {\bibinfo {volume} {56}},\ \bibinfo {pages} {095010}
  (\bibinfo {year} {2014})}\BibitemShut {NoStop}%
\bibitem [{\citenamefont {Horton}(2000)}]{horton2000h}%
  \BibitemOpen
  \bibfield  {author} {\bibinfo {author} {\bibfnamefont {L.}~\bibnamefont
  {Horton}},\ }\href@noop {} {\bibfield  {journal} {\bibinfo  {journal} {Plasma
  Physics and Controlled Fusion}\ }\textbf {\bibinfo {volume} {42}},\ \bibinfo
  {pages} {A37} (\bibinfo {year} {2000})}\BibitemShut {NoStop}%
\bibitem [{\citenamefont {Doyle}\ \emph {et~al.}(2007)\citenamefont {Doyle},
  \citenamefont {Houlberg}, \citenamefont {Kamada}, \citenamefont {Mukhovatov},
  \citenamefont {Osborne}, \citenamefont {Polevoi}, \citenamefont {Bateman},
  \citenamefont {Connor}, \citenamefont {Cordey}, \citenamefont {Fujita} \emph
  {et~al.}}]{doyle2007plasma}%
  \BibitemOpen
  \bibfield  {author} {\bibinfo {author} {\bibfnamefont {E.}~\bibnamefont
  {Doyle}}, \bibinfo {author} {\bibfnamefont {W.}~\bibnamefont {Houlberg}},
  \bibinfo {author} {\bibfnamefont {Y.}~\bibnamefont {Kamada}}, \bibinfo
  {author} {\bibfnamefont {V.}~\bibnamefont {Mukhovatov}}, \bibinfo {author}
  {\bibfnamefont {T.}~\bibnamefont {Osborne}}, \bibinfo {author} {\bibfnamefont
  {A.}~\bibnamefont {Polevoi}}, \bibinfo {author} {\bibfnamefont
  {G.}~\bibnamefont {Bateman}}, \bibinfo {author} {\bibfnamefont
  {J.}~\bibnamefont {Connor}}, \bibinfo {author} {\bibfnamefont
  {J.}~\bibnamefont {Cordey}}, \bibinfo {author} {\bibfnamefont
  {T.}~\bibnamefont {Fujita}},  \emph {et~al.},\ }\href@noop {} {\bibfield
  {journal} {\bibinfo  {journal} {Nuclear Fusion}\ }\textbf {\bibinfo {volume}
  {47}},\ \bibinfo {pages} {S18} (\bibinfo {year} {2007})}\BibitemShut
  {NoStop}%
\bibitem [{\citenamefont {Connor}\ and\ \citenamefont
  {Wilson}(2000)}]{connor2000review}%
  \BibitemOpen
  \bibfield  {author} {\bibinfo {author} {\bibfnamefont {J.}~\bibnamefont
  {Connor}}\ and\ \bibinfo {author} {\bibfnamefont {H.}~\bibnamefont
  {Wilson}},\ }\href@noop {} {\bibfield  {journal} {\bibinfo  {journal} {Plasma
  Physics and Controlled Fusion}\ }\textbf {\bibinfo {volume} {42}},\ \bibinfo
  {pages} {R1} (\bibinfo {year} {2000})}\BibitemShut {NoStop}%
\bibitem [{\citenamefont {Daughton}\ \emph {et~al.}(1998)\citenamefont
  {Daughton}, \citenamefont {Catto}, \citenamefont {Coppi},\ and\ \citenamefont
  {Krasheninnikov}}]{daughton1998interchange}%
  \BibitemOpen
  \bibfield  {author} {\bibinfo {author} {\bibfnamefont {W.}~\bibnamefont
  {Daughton}}, \bibinfo {author} {\bibfnamefont {P.~J.}\ \bibnamefont {Catto}},
  \bibinfo {author} {\bibfnamefont {B.}~\bibnamefont {Coppi}}, \ and\ \bibinfo
  {author} {\bibfnamefont {S.}~\bibnamefont {Krasheninnikov}},\ }\href@noop {}
  {\bibfield  {journal} {\bibinfo  {journal} {Physics of Plasmas}\ }\textbf
  {\bibinfo {volume} {5}},\ \bibinfo {pages} {2217} (\bibinfo {year}
  {1998})}\BibitemShut {NoStop}%
\bibitem [{\citenamefont {{\"O}dblom}\ \emph {et~al.}(1999)\citenamefont
  {{\"O}dblom}, \citenamefont {Catto},\ and\ \citenamefont
  {Krasheninnikov}}]{odblom1999neutrals}%
  \BibitemOpen
  \bibfield  {author} {\bibinfo {author} {\bibfnamefont {A.}~\bibnamefont
  {{\"O}dblom}}, \bibinfo {author} {\bibfnamefont {P.}~\bibnamefont {Catto}}, \
  and\ \bibinfo {author} {\bibfnamefont {S.}~\bibnamefont {Krasheninnikov}},\
  }\href@noop {} {\bibfield  {journal} {\bibinfo  {journal} {Physics of
  Plasmas}\ }\textbf {\bibinfo {volume} {6}},\ \bibinfo {pages} {3239}
  (\bibinfo {year} {1999})}\BibitemShut {NoStop}%
\bibitem [{\citenamefont {Monier-Garbet}\ \emph {et~al.}(1997)\citenamefont
  {Monier-Garbet}, \citenamefont {Burrell}, \citenamefont {Hinton},
  \citenamefont {Kim}, \citenamefont {Garbet},\ and\ \citenamefont
  {Groebner}}]{monier1997effects}%
  \BibitemOpen
  \bibfield  {author} {\bibinfo {author} {\bibfnamefont {P.}~\bibnamefont
  {Monier-Garbet}}, \bibinfo {author} {\bibfnamefont {K.}~\bibnamefont
  {Burrell}}, \bibinfo {author} {\bibfnamefont {F.}~\bibnamefont {Hinton}},
  \bibinfo {author} {\bibfnamefont {J.}~\bibnamefont {Kim}}, \bibinfo {author}
  {\bibfnamefont {X.}~\bibnamefont {Garbet}}, \ and\ \bibinfo {author}
  {\bibfnamefont {R.}~\bibnamefont {Groebner}},\ }\href@noop {} {\bibfield
  {journal} {\bibinfo  {journal} {Nuclear fusion}\ }\textbf {\bibinfo {volume}
  {37}},\ \bibinfo {pages} {403} (\bibinfo {year} {1997})}\BibitemShut
  {NoStop}%
\bibitem [{\citenamefont {F{\"u}l{\"o}p}\ \emph {et~al.}(2001)\citenamefont
  {F{\"u}l{\"o}p}, \citenamefont {Catto},\ and\ \citenamefont
  {Helander}}]{fulop2001effect}%
  \BibitemOpen
  \bibfield  {author} {\bibinfo {author} {\bibfnamefont {T.}~\bibnamefont
  {F{\"u}l{\"o}p}}, \bibinfo {author} {\bibfnamefont {P.~J.}\ \bibnamefont
  {Catto}}, \ and\ \bibinfo {author} {\bibfnamefont {P.}~\bibnamefont
  {Helander}},\ }\href@noop {} {\bibfield  {journal} {\bibinfo  {journal}
  {Physics of Plasmas}\ }\textbf {\bibinfo {volume} {8}},\ \bibinfo {pages}
  {5214} (\bibinfo {year} {2001})}\BibitemShut {NoStop}%
\bibitem [{\citenamefont {D’Ippolito}\ and\ \citenamefont
  {Myra}(2002)}]{d2002effect}%
  \BibitemOpen
  \bibfield  {author} {\bibinfo {author} {\bibfnamefont {D.}~\bibnamefont
  {D’Ippolito}}\ and\ \bibinfo {author} {\bibfnamefont {J.}~\bibnamefont
  {Myra}},\ }\href@noop {} {\bibfield  {journal} {\bibinfo  {journal} {Physics
  of Plasmas}\ }\textbf {\bibinfo {volume} {9}},\ \bibinfo {pages} {853}
  (\bibinfo {year} {2002})}\BibitemShut {NoStop}%
\bibitem [{\citenamefont {Wersal}\ and\ \citenamefont
  {Ricci}(2015)}]{wersal2015first}%
  \BibitemOpen
  \bibfield  {author} {\bibinfo {author} {\bibfnamefont {C.}~\bibnamefont
  {Wersal}}\ and\ \bibinfo {author} {\bibfnamefont {P.}~\bibnamefont {Ricci}},\
  }\href@noop {} {\bibfield  {journal} {\bibinfo  {journal} {Nuclear Fusion}\
  }\textbf {\bibinfo {volume} {55}},\ \bibinfo {pages} {123014} (\bibinfo
  {year} {2015})}\BibitemShut {NoStop}%
\bibitem [{\citenamefont {Stotler}\ \emph {et~al.}(2017)\citenamefont
  {Stotler}, \citenamefont {Lang}, \citenamefont {Chang}, \citenamefont
  {Churchill},\ and\ \citenamefont {Ku}}]{stotler2017neutral}%
  \BibitemOpen
  \bibfield  {author} {\bibinfo {author} {\bibfnamefont {D.}~\bibnamefont
  {Stotler}}, \bibinfo {author} {\bibfnamefont {J.}~\bibnamefont {Lang}},
  \bibinfo {author} {\bibfnamefont {C.}~\bibnamefont {Chang}}, \bibinfo
  {author} {\bibfnamefont {R.}~\bibnamefont {Churchill}}, \ and\ \bibinfo
  {author} {\bibfnamefont {S.}~\bibnamefont {Ku}},\ }\href@noop {} {\bibfield
  {journal} {\bibinfo  {journal} {Nuclear Fusion}\ }\textbf {\bibinfo {volume}
  {57}},\ \bibinfo {pages} {086028} (\bibinfo {year} {2017})}\BibitemShut
  {NoStop}%
\bibitem [{\citenamefont {Thrys{\o}e}\ \emph {et~al.}(2018)\citenamefont
  {Thrys{\o}e}, \citenamefont {L{\o}iten}, \citenamefont {Madsen},
  \citenamefont {Naulin}, \citenamefont {Nielsen},\ and\ \citenamefont
  {Rasmussen}}]{thrysoe2018plasma}%
  \BibitemOpen
  \bibfield  {author} {\bibinfo {author} {\bibfnamefont {A.~S.}\ \bibnamefont
  {Thrys{\o}e}}, \bibinfo {author} {\bibfnamefont {M.}~\bibnamefont
  {L{\o}iten}}, \bibinfo {author} {\bibfnamefont {J.}~\bibnamefont {Madsen}},
  \bibinfo {author} {\bibfnamefont {V.}~\bibnamefont {Naulin}}, \bibinfo
  {author} {\bibfnamefont {A.}~\bibnamefont {Nielsen}}, \ and\ \bibinfo
  {author} {\bibfnamefont {J.~J.}\ \bibnamefont {Rasmussen}},\ }\href@noop {}
  {\bibfield  {journal} {\bibinfo  {journal} {Physics of Plasmas}\ }\textbf
  {\bibinfo {volume} {25}},\ \bibinfo {pages} {032307} (\bibinfo {year}
  {2018})}\BibitemShut {NoStop}%
\bibitem [{\citenamefont {Bisai}\ and\ \citenamefont
  {Kaw}(2018)}]{bisai2018influence}%
  \BibitemOpen
  \bibfield  {author} {\bibinfo {author} {\bibfnamefont {N.}~\bibnamefont
  {Bisai}}\ and\ \bibinfo {author} {\bibfnamefont {P.}~\bibnamefont {Kaw}},\
  }\href@noop {} {\bibfield  {journal} {\bibinfo  {journal} {Physics of
  Plasmas}\ }\textbf {\bibinfo {volume} {25}},\ \bibinfo {pages} {012503}
  (\bibinfo {year} {2018})}\BibitemShut {NoStop}%
\bibitem [{\citenamefont {Helander}\ \emph {et~al.}(1994)\citenamefont
  {Helander}, \citenamefont {Krasheninnikov},\ and\ \citenamefont
  {Catto}}]{helander1994fluid}%
  \BibitemOpen
  \bibfield  {author} {\bibinfo {author} {\bibfnamefont {P.}~\bibnamefont
  {Helander}}, \bibinfo {author} {\bibfnamefont {S.}~\bibnamefont
  {Krasheninnikov}}, \ and\ \bibinfo {author} {\bibfnamefont {P.}~\bibnamefont
  {Catto}},\ }\href@noop {} {\bibfield  {journal} {\bibinfo  {journal} {Physics
  of plasmas}\ }\textbf {\bibinfo {volume} {1}},\ \bibinfo {pages} {3174}
  (\bibinfo {year} {1994})}\BibitemShut {NoStop}%
\bibitem [{\citenamefont {Diamond}\ \emph {et~al.}(2005)\citenamefont
  {Diamond}, \citenamefont {Itoh}, \citenamefont {Itoh},\ and\ \citenamefont
  {Hahm}}]{diamond2005zonal}%
  \BibitemOpen
  \bibfield  {author} {\bibinfo {author} {\bibfnamefont {P.~H.}\ \bibnamefont
  {Diamond}}, \bibinfo {author} {\bibfnamefont {S.}~\bibnamefont {Itoh}},
  \bibinfo {author} {\bibfnamefont {K.}~\bibnamefont {Itoh}}, \ and\ \bibinfo
  {author} {\bibfnamefont {T.}~\bibnamefont {Hahm}},\ }\href@noop {} {\bibfield
   {journal} {\bibinfo  {journal} {Plasma Physics and Controlled Fusion}\
  }\textbf {\bibinfo {volume} {47}},\ \bibinfo {pages} {R35} (\bibinfo {year}
  {2005})}\BibitemShut {NoStop}%
\bibitem [{\citenamefont {Fujisawa}(2008)}]{fujisawa2008review}%
  \BibitemOpen
  \bibfield  {author} {\bibinfo {author} {\bibfnamefont {A.}~\bibnamefont
  {Fujisawa}},\ }\href@noop {} {\bibfield  {journal} {\bibinfo  {journal}
  {Nuclear Fusion}\ }\textbf {\bibinfo {volume} {49}},\ \bibinfo {pages}
  {013001} (\bibinfo {year} {2008})}\BibitemShut {NoStop}%
\bibitem [{\citenamefont {Numata}\ \emph {et~al.}(2007)\citenamefont {Numata},
  \citenamefont {Ball},\ and\ \citenamefont {Dewar}}]{numata2007bifurcation}%
  \BibitemOpen
  \bibfield  {author} {\bibinfo {author} {\bibfnamefont {R.}~\bibnamefont
  {Numata}}, \bibinfo {author} {\bibfnamefont {R.}~\bibnamefont {Ball}}, \ and\
  \bibinfo {author} {\bibfnamefont {R.~L.}\ \bibnamefont {Dewar}},\ }\href@noop
  {} {\bibfield  {journal} {\bibinfo  {journal} {Physics of Plasmas}\ }\textbf
  {\bibinfo {volume} {14}},\ \bibinfo {pages} {102312} (\bibinfo {year}
  {2007})}\BibitemShut {NoStop}%
\bibitem [{\citenamefont {Hasegawa}\ and\ \citenamefont
  {Wakatani}(1983)}]{hasegawa1983plasma}%
  \BibitemOpen
  \bibfield  {author} {\bibinfo {author} {\bibfnamefont {A.}~\bibnamefont
  {Hasegawa}}\ and\ \bibinfo {author} {\bibfnamefont {M.}~\bibnamefont
  {Wakatani}},\ }\href@noop {} {\bibfield  {journal} {\bibinfo  {journal}
  {Physical Review Letters}\ }\textbf {\bibinfo {volume} {50}},\ \bibinfo
  {pages} {682} (\bibinfo {year} {1983})}\BibitemShut {NoStop}%
\bibitem [{\citenamefont {Sun}\ \emph {et~al.}(2015)\citenamefont {Sun},
  \citenamefont {Wolfrum}, \citenamefont {Eich}, \citenamefont {Kurzan},
  \citenamefont {Potzel}, \citenamefont {Stroth} \emph
  {et~al.}}]{sun2015study}%
  \BibitemOpen
  \bibfield  {author} {\bibinfo {author} {\bibfnamefont {H.}~\bibnamefont
  {Sun}}, \bibinfo {author} {\bibfnamefont {E.}~\bibnamefont {Wolfrum}},
  \bibinfo {author} {\bibfnamefont {T.}~\bibnamefont {Eich}}, \bibinfo {author}
  {\bibfnamefont {B.}~\bibnamefont {Kurzan}}, \bibinfo {author} {\bibfnamefont
  {S.}~\bibnamefont {Potzel}}, \bibinfo {author} {\bibfnamefont
  {U.}~\bibnamefont {Stroth}},  \emph {et~al.},\ }\href@noop {} {\bibfield
  {journal} {\bibinfo  {journal} {Plasma Physics and Controlled Fusion}\
  }\textbf {\bibinfo {volume} {57}},\ \bibinfo {pages} {125011} (\bibinfo
  {year} {2015})}\BibitemShut {NoStop}%
\bibitem [{\citenamefont {Rognlien}\ \emph {et~al.}(1992)\citenamefont
  {Rognlien}, \citenamefont {Milovich}, \citenamefont {Rensink},\ and\
  \citenamefont {Porter}}]{rognlien1992fully}%
  \BibitemOpen
  \bibfield  {author} {\bibinfo {author} {\bibfnamefont {T.~D.}\ \bibnamefont
  {Rognlien}}, \bibinfo {author} {\bibfnamefont {J.}~\bibnamefont {Milovich}},
  \bibinfo {author} {\bibfnamefont {M.}~\bibnamefont {Rensink}}, \ and\
  \bibinfo {author} {\bibfnamefont {G.}~\bibnamefont {Porter}},\ }\href@noop {}
  {\bibfield  {journal} {\bibinfo  {journal} {Journal of nuclear materials}\
  }\textbf {\bibinfo {volume} {196}},\ \bibinfo {pages} {347} (\bibinfo {year}
  {1992})}\BibitemShut {NoStop}%
\bibitem [{\citenamefont {Hasegawa}\ and\ \citenamefont
  {Mima}(1978)}]{hasegawa1978pseudo}%
  \BibitemOpen
  \bibfield  {author} {\bibinfo {author} {\bibfnamefont {A.}~\bibnamefont
  {Hasegawa}}\ and\ \bibinfo {author} {\bibfnamefont {K.}~\bibnamefont
  {Mima}},\ }\href@noop {} {\bibfield  {journal} {\bibinfo  {journal} {The
  Physics of Fluids}\ }\textbf {\bibinfo {volume} {21}},\ \bibinfo {pages} {87}
  (\bibinfo {year} {1978})}\BibitemShut {NoStop}%
\bibitem [{\citenamefont {Smolyakov}\ \emph
  {et~al.}(2000{\natexlab{a}})\citenamefont {Smolyakov}, \citenamefont
  {Diamond},\ and\ \citenamefont {Malkov}}]{smolyakov2000coherent}%
  \BibitemOpen
  \bibfield  {author} {\bibinfo {author} {\bibfnamefont {A.}~\bibnamefont
  {Smolyakov}}, \bibinfo {author} {\bibfnamefont {P.}~\bibnamefont {Diamond}},
  \ and\ \bibinfo {author} {\bibfnamefont {M.}~\bibnamefont {Malkov}},\
  }\href@noop {} {\bibfield  {journal} {\bibinfo  {journal} {Physical review
  letters}\ }\textbf {\bibinfo {volume} {84}},\ \bibinfo {pages} {491}
  (\bibinfo {year} {2000}{\natexlab{a}})}\BibitemShut {NoStop}%
\bibitem [{\citenamefont {Smolyakov}\ \emph
  {et~al.}(2000{\natexlab{b}})\citenamefont {Smolyakov}, \citenamefont
  {Diamond},\ and\ \citenamefont {Shevchenko}}]{smolyakov2000zonal}%
  \BibitemOpen
  \bibfield  {author} {\bibinfo {author} {\bibfnamefont {A.}~\bibnamefont
  {Smolyakov}}, \bibinfo {author} {\bibfnamefont {P.}~\bibnamefont {Diamond}},
  \ and\ \bibinfo {author} {\bibfnamefont {V.}~\bibnamefont {Shevchenko}},\
  }\href@noop {} {\bibfield  {journal} {\bibinfo  {journal} {Physics of
  Plasmas}\ }\textbf {\bibinfo {volume} {7}},\ \bibinfo {pages} {1349}
  (\bibinfo {year} {2000}{\natexlab{b}})}\BibitemShut {NoStop}%
\bibitem [{\citenamefont {Braginskii}(1965)}]{braginskii1965transport}%
  \BibitemOpen
  \bibfield  {author} {\bibinfo {author} {\bibfnamefont {S.}~\bibnamefont
  {Braginskii}},\ }\href@noop {} {\bibfield  {journal} {\bibinfo  {journal}
  {Reviews of plasma physics}\ }\textbf {\bibinfo {volume} {1}} (\bibinfo
  {year} {1965})}\BibitemShut {NoStop}%
\bibitem [{\citenamefont {Burns}\ \emph {et~al.}(2019)\citenamefont {Burns},
  \citenamefont {Vasil}, \citenamefont {Oishi}, \citenamefont {Lecoanet},\ and\
  \citenamefont {Brown}}]{2019arXiv190510388B}%
  \BibitemOpen
  \bibfield  {author} {\bibinfo {author} {\bibfnamefont {K.~J.}\ \bibnamefont
  {Burns}}, \bibinfo {author} {\bibfnamefont {G.~M.}\ \bibnamefont {Vasil}},
  \bibinfo {author} {\bibfnamefont {J.~S.}\ \bibnamefont {Oishi}}, \bibinfo
  {author} {\bibfnamefont {D.}~\bibnamefont {Lecoanet}}, \ and\ \bibinfo
  {author} {\bibfnamefont {B.~P.}\ \bibnamefont {Brown}},\ }\href@noop {}
  {\enquote {\bibinfo {title} {Dedalus: A flexible framework for numerical
  simulations with spectral methods},}\ } (\bibinfo {year} {2019}),\ \Eprint
  {http://arxiv.org/abs/1905.10388.} {arXiv:1905.10388.} \BibitemShut {NoStop}%
~http://dedalus-project.org
\bibitem [{\citenamefont {Diamond}\ \emph {et~al.}(1994)\citenamefont
  {Diamond}, \citenamefont {Liang}, \citenamefont {Carreras},\ and\
  \citenamefont {Terry}}]{diamond1994self}%
  \BibitemOpen
  \bibfield  {author} {\bibinfo {author} {\bibfnamefont {P.}~\bibnamefont
  {Diamond}}, \bibinfo {author} {\bibfnamefont {Y.-M.}\ \bibnamefont {Liang}},
  \bibinfo {author} {\bibfnamefont {B.}~\bibnamefont {Carreras}}, \ and\
  \bibinfo {author} {\bibfnamefont {P.}~\bibnamefont {Terry}},\ }\href@noop {}
  {\bibfield  {journal} {\bibinfo  {journal} {Physical review letters}\
  }\textbf {\bibinfo {volume} {72}},\ \bibinfo {pages} {2565} (\bibinfo {year}
  {1994})}\BibitemShut {NoStop}%
\bibitem [{\citenamefont {Masline}\ \emph {et~al.}(2019)\citenamefont
  {Masline}, \citenamefont {Smirnov},\ and\ \citenamefont
  {Krasheninnikov}}]{masline2019influence}%
  \BibitemOpen
  \bibfield  {author} {\bibinfo {author} {\bibfnamefont {R.}~\bibnamefont
  {Masline}}, \bibinfo {author} {\bibfnamefont {R.~D.}\ \bibnamefont
  {Smirnov}}, \ and\ \bibinfo {author} {\bibfnamefont {S.~I.}\ \bibnamefont
  {Krasheninnikov}},\ }\href@noop {} {\bibfield  {journal} {\bibinfo  {journal}
  {Contributions to Plasma Physics}\ ,\ \bibinfo {pages} {e201900097}}
  (\bibinfo {year} {2019})}\BibitemShut {NoStop}%
\bibitem [{\citenamefont {Krasheninnikov}\ \emph {et~al.}(2017)\citenamefont
  {Krasheninnikov}, \citenamefont {Kukushkin}, \citenamefont {Pshenov},
  \citenamefont {Smolyakov},\ and\ \citenamefont
  {Zhang}}]{krasheninnikov2017stability}%
  \BibitemOpen
  \bibfield  {author} {\bibinfo {author} {\bibfnamefont {S.}~\bibnamefont
  {Krasheninnikov}}, \bibinfo {author} {\bibfnamefont {A.}~\bibnamefont
  {Kukushkin}}, \bibinfo {author} {\bibfnamefont {A.}~\bibnamefont {Pshenov}},
  \bibinfo {author} {\bibfnamefont {A.}~\bibnamefont {Smolyakov}}, \ and\
  \bibinfo {author} {\bibfnamefont {Y.}~\bibnamefont {Zhang}},\ }\href@noop {}
  {\bibfield  {journal} {\bibinfo  {journal} {Nuclear Materials and Energy}\
  }\textbf {\bibinfo {volume} {12}},\ \bibinfo {pages} {1061} (\bibinfo {year}
  {2017})}\BibitemShut {NoStop}%
\bibitem [{\citenamefont {Jaervinen}\ \emph {et~al.}(2018)\citenamefont
  {Jaervinen}, \citenamefont {Allen}, \citenamefont {Eldon}, \citenamefont
  {Fenstermacher}, \citenamefont {Groth}, \citenamefont {Hill}, \citenamefont
  {Leonard}, \citenamefont {McLean}, \citenamefont {Porter}, \citenamefont
  {Rognlien} \emph {et~al.}}]{jaervinen2018b}%
  \BibitemOpen
  \bibfield  {author} {\bibinfo {author} {\bibfnamefont {A.}~\bibnamefont
  {Jaervinen}}, \bibinfo {author} {\bibfnamefont {S.}~\bibnamefont {Allen}},
  \bibinfo {author} {\bibfnamefont {D.}~\bibnamefont {Eldon}}, \bibinfo
  {author} {\bibfnamefont {M.}~\bibnamefont {Fenstermacher}}, \bibinfo {author}
  {\bibfnamefont {M.}~\bibnamefont {Groth}}, \bibinfo {author} {\bibfnamefont
  {D.~N.}\ \bibnamefont {Hill}}, \bibinfo {author} {\bibfnamefont
  {A.}~\bibnamefont {Leonard}}, \bibinfo {author} {\bibfnamefont
  {A.}~\bibnamefont {McLean}}, \bibinfo {author} {\bibfnamefont
  {G.}~\bibnamefont {Porter}}, \bibinfo {author} {\bibfnamefont
  {T.}~\bibnamefont {Rognlien}},  \emph {et~al.},\ }\href@noop {} {\bibfield
  {journal} {\bibinfo  {journal} {Physical review letters}\ }\textbf {\bibinfo
  {volume} {121}},\ \bibinfo {pages} {075001} (\bibinfo {year}
  {2018})}\BibitemShut {NoStop}%
\end{thebibliography}%

\end{document}